\documentclass{IOS-Book-Article}

\usepackage{times}
\normalfont

\newcommand{\be}{\begin{equation}}
\newcommand{\ee}{\end{equation}}
\newcommand{\ba}{\begin{eqnarray}}
\newcommand{\ea}{\end{eqnarray}}
\newcommand{\ban}{\begin{eqnarray*}}
\newcommand{\ean}{\end{eqnarray*}}

\newcommand{\ket}[1]{\mbox{$ | #1 \rangle $}}
\newcommand{\bra}[1]{\mbox{$ \langle #1 | $}}

\newcommand{\demi}{\frac{1}{2}}

\newcommand{\one}{\leavevmode\hbox{\small1\normalsize\kern-.33em1}}

\begin{document}
\begin{frontmatter}                           

\title{QKD: a million signal task}
\runningtitle{QKD: a million signal task}

\author[A]{\fnms{Valerio} \snm{Scarani}%
\thanks{Corresponding Author: Centre for Quantum Technologies, 3 Science Drive 2, Singapore 117543. E-mail: physv@nus.edu.sg.}}
\runningauthor{V. Scarani}
\address[A]{Centre for Quantum Technologies and Department of Physics, National University of Singapore, Singapore.}

\begin{abstract}
I review the ideas and main results in the derivation of security bounds in quantum key distribution for keys of finite length. In particular, all the detailed studies on specific protocols and implementations indicate that no secret key can be extracted if the number of processed signals per run is smaller than $10^5-10^6$. I show how these numbers can be recovered from very basic estimates.
\end{abstract}

\end{frontmatter}

\thispagestyle{empty}
\pagestyle{empty}

\section{Introduction}

In the very first proof of unconditional security of quantum key distribution (QKD), Mayers already stressed the need to obtain \textit{finite-key bounds} and paved the path for this task \cite{mayers}. The complex development of QKD, which has been reviewed elsewhere \cite{review}, explains why this task was finally achieved only recently.

This text is a concise presentation of finite-key security proofs, written for readers that are already familiar with the main notions of QKD. I use the approach developed by Renato Renner in his thesis \cite{rennerthesis} and applied later to the specific study of finite-key effects \cite{scaren1, scaren2, caisca}. Independently, and actually some months earlier than Renner and myself, Hayashi has also developed finite-key analysis \cite{hay2,hay3,hay4}. A recent work by Fung, Ma and Chau tackles the problem from yet another approach \cite{fung}. It must be stressed that all these approaches are recognized to be ultimately based on the same definition of security and lead to the same order of magnitude for the finite-key corrections, although there may be differences in details.

\section{Finite-key corrections}

In a QKD protocol, Alice and Bob have to infer the information that could have leaked to the eavesdropper Eve on the basis of some measured parameters: error rates, detection rates, and others. These parameters and the way of measuring them define the protocol. Let us refer to them as $\mathbf{V}$ measured value of the parameters.

Let $N$ be total number of signals exchanged by Alice and Bob in a run of key exchange, $n$ the number of signals kept for the key (i.e. the length of the raw key) and $m$ the number of signals used for parameter estimation. The length of the secret key to be extracted is denoted by $\ell$. The usually quoted number is the \textit{secret key fraction} $r=\frac{\ell}{N}$.

It is important to note from the start that a \textit{run of key exchange} is defined as the number of signals that shall be processed together: obviously, the duration of the key exchange cannot increase its security. In all this text, I focus on one-way post-processing without pre-processing, so specifically a run is defined by the size of the blocks on which privacy amplification is performed.

In the asymptotic limit, the expression of $r$ is by now well-known:
\ba
r_{\infty}&=&\lim_{N\rightarrow\infty}\frac{\ell}{N}\,=\,\min_{E|\mathbf{V}}H(A|E)-H(A|B)\label{rinf}
\ea
It has an intuitive meaning: the fraction that can be extracted is the uncertainty of Eve minus the uncertainty of Bob (or equivalently, the information of Bob minus the information of Eve) on Alice's key.

The finite-key bound, to be proved below, reads
\ba
r_{N}&=&\frac{\ell}{N}\,=\,\frac{n}{N}\left[\min_{E|\mathbf{V\pm\Delta V}}H(A|E)-\Delta(n)-\textrm{leak}_{EC}/n\right]\,.\label{rfinite}
\label{finitebound}\ea
Four differences between (\ref{rinf}) and (\ref{rfinite}) are worth stressing:
\begin{enumerate}
\item $\frac{n}{N}$: in the asymptotic case, the protocol can be adapted so that almost all the signals are used for the raw key and only a negligible fraction is used for parameter estimation \cite{LoChauArdehali}. This is of course no longer the case in the finite-key regime: one needs to devote enough many signals to parameter estimation in order to have good statistics.
\item $V\pm\Delta V$: the statistical estimates of the parameters come with fluctuations for finite sampling.
\item $\textrm{leak}_{EC}$: this is the information that leaks out to Eve during error correction; in the asymptotic case, it is given by the Shannon bound $H(A|B)$, but for practical codes on finite samples the Shannon bound may not be reached.
\item $\Delta(n)$: while the three previous items are pretty obvious and and had in fact been anticipated in many works, both theoretical and experimental, this term is the challenging one. It comes as an overhead to privacy amplification: the task itself cannot be carried out perfectly on finite samples.
\end{enumerate}
In summary, the theoretical challenge of finite-key analysis mainly consists of obtaining an expression for $\Delta(n)$. As we shall show, once this expression is obtained, one can study which of the corrections has the largest effect; it will turn out that the most dramatic correction comes from the ``obvious'' need to take into account $\Delta V$, the statistical fluctuations in the estimate of the parameters.

\section{Derivation of the finite-key bound}

\subsection{Scenario}
The following equation summarizes the steps of a QKD exchange:
\ban
\Phi_{AB}^{\otimes N}\,\stackrel{\mathrm{Eve}}{\longrightarrow}\, \Psi_{A^NB^NE}&\stackrel{\mathrm{Meas.}}{\longrightarrow}&\sum_{\underline{a},\underline{b}\in\{0,1\}^n} p(\underline{a},\underline{b})\ket{\underline{a}}\bra{\underline{a}}\otimes \ket{\underline{b}}\bra{\underline{b}}\otimes \rho_E^{\underline{a},\underline{b}}\;\textrm{ and parameters }\mathbf{V}\\
& \stackrel{\mathrm{EC}}{\longrightarrow}& \sum_{\underline{a}\in\{0,1\}^n} p(\underline{a})\ket{\underline{a}}\bra{\underline{a}}\otimes \ket{\underline{a}}\bra{\underline{a}}\otimes \rho_{{\cal E}}^{\underline{a}}\\
& \stackrel{\mathrm{PA(\mathbf{V})}}{\longrightarrow}& \Big(\sum_{\underline{k}\in\{0,1\}^\ell} \frac{1}{2^\ell}\ket{\underline{k}}\bra{\underline{k}}\otimes \ket{\underline{k}}\bra{\underline{k}}\Big)\otimes \rho_{{\cal E}}
\,\,\equiv \rho_U\otimes\rho_{{\cal E}}\,.\ean
In words: on the quantum channel, Eve interacts with the $N$ signals flying between Alice and Bob; by giving Eve a purification of the whole state, we give her the maximal possible information. When Alice and Bob measure, the total state becomes a classical-classical-quantum (ccq) state. After error correction (EC), Alice's and Bob's classical lists are equal, but Eve has gained some additional classical information: $\rho_{{\cal E}}^{\underline{a}}=\sum_{\underline{b}}p(\underline{b}|\underline{a})\rho_E^{\underline{a},\underline{b}}\otimes\ket{C(\underline{b}\rightarrow\underline{a})}\bra{C(\underline{b}\rightarrow\underline{a})}$. After privacy amplification (PA), the lists of Alice and Bob are shorter, but still equal and now drawn from a completely random distribution, on which Eve has no information.

This is for the ideal case. In order to estimate $\ell$, the first step consists in defining a security parameter. The most convenient choice, actually the only one with the suitable properties known to date (see \cite{scaren1} for a discussion), is the \textit{probability that the processed state differs from the ideal one}:
\ba
\epsilon&=&\demi\,\textrm{Tr}\left|\rho_{KE}-\rho_U\otimes\rho_E\right|\,.
\ea
So, now we choose a value for $\epsilon$ (say $10^{-9}$, i.e. I want at most one run in one billion to go wrong), our data processing defines $N$: how to get the corresponding $\ell$? This is the subject of the next paragraph.

\subsection{Renner's version of Murphy's law}

The recipe is: quantify everything that can go wrong, going backwards, i.e. starting from PA and going up to parameter estimation. Here is what it gives in detail:

\begin{enumerate}

\item First step: \textbf{Privacy amplification}. The probability that PA fails even if everything has been perfectly carried out is
\ba
\epsilon_{PA}&=& 2^{-\demi\left(H_{min}^{\bar{\epsilon}}(A^n|{\cal E})-\ell\right)}\,.\label{epa}\ea
As usual, for the proof and the exact definitions I refer to the original papers quoted above. Let us however get an intuition of this result. The symbol ${\cal E}$ represent all that Eve has learned, both from attacking the quantum channel and from listening to error correction. Alice's information is denoted by $A^n$ to remind that it is a sequence of $n$ bits. The main object introduced here is $H_{min}^{\bar{\epsilon}}(.|.)$ is the quantum conditional \textit{smooth min-entropy}, with ${\bar{\epsilon}}$ the smoothing parameter. It is not astonishing that a min-entropy appears here: as we said above, the final key should be randomly distributed given Eve's knowledge. In classical information theory, the min-entropy precisely quantifies the fraction of completely random bits one can extract from a given list of partially random bits. The challenge consisted in defining the quantity that plays the same role, in the case where an adversary has quantum information. This was done in Renner's thesis \cite{rennerthesis}. So, in a sense, (\ref{epa}) can be seen as a definition of smooth min-entropy, or better, as one of the desiderata: the smooth min-entropy must be such that (\ref{epa}) holds.

Now, for the right-hand side of (\ref{epa}) to be bounded, one must have
\ba
\ell\,\leq\, H_{min}(A^n|{\cal E})-2\log\frac{1}{\epsilon_{PA}}\,.
\ea
It is crucial to understand that this is \textit{already} the desired bound for $\ell$. Unfortunately, there is no easy way of computing $H_{min}(A^n|{\cal E})$: the next steps will be merely bound on this last quantity, reaching to an ultimately computable expression.

\item Second step: \textbf{error correction}. First, we have to add an $\epsilon_{EC}$ to the failure probability $\epsilon$. This measures the probability that the EC procedure was apparently successful but some uncorrected errors remain\footnote{This may sound cumbersome at first reading, but it is actually clear. Any EC procedure has a possible red-flag outcome: ``Sorry, the procedure was not successful''. In this case, one just discards the raw key, without compromising the security. Problems arise only when there are still uncorrected errors left but the red flag is not raised.}. Moreover, as we said, during EC, additional data $C$ has been provided to Eve, giving an amount of information $\textrm{leak}_{EC}$; by using a chain rule, one can split this information from the one acquired in attacking the quantum channel, denoted by $E$. So we obtain the expression
\ba
H_{min}(A^n|{\cal E})&=&H_{min}(A^n|E)-\textrm{leak}_{EC}\,.
\ea
Theory predicts $\textrm{leak}_{EC}\approx fH_0(A^n|B^n)+\log\frac{2}{\epsilon_{EC}}$, but ultimately this is number of bits \textit{really} exchanged during EC: it does not need theoretical modeling (unless one is working on designing a better EC code), just pick it from the real run of the EC code.

\item Third step: \textbf{Eve's attack on the quantum channel}. Now we are back at the level of quantum interaction and we have to compute $H_{min}(A^n|E)$: how much information Eve might have acquired, given the observed disturbances $\textbf{V}$. Under the assumption of collective attacks, one finds
\ba
H_{min}(A^n|E=E^n)&\leq& n\left[H(A|E)-7\sqrt{\frac{\log(2/\bar{\epsilon})}{n}}\right]\,.
\ea
The assumption of collective attacks is not restrictive for the Bennett-Brassard 1984 (BB84) protocol and many others, under the validity of the squashing condition\footnote{This condition means that one can prove the security of a protocol on the level of qubits, even if physically light fields are infinitely dimensional systems. In other words, one can ``squash'' the meaningful degrees of freedom down to qubits. This statement has been rigorously proved for the BB84 protocol \cite{bea08,tsu08}.}. In order to remove this assumption, one could resort to the suitable De Finetti theorem, but the bound scales very badly. A much more promising approach is based on Ref.~\cite{postsel}, but to my knowledge it has not been applied yet in the context of finite-keys.

\item Fourth step: \textbf{Optimize over Eve's attacks}. Since we don't know which attack Eve has actually performed, $H(A|E)$ must be computed for the best possible attack compatible with the measured parameters $\mathbf{V}$. But these parameters have been measured on a finite sample: they are subject to fluctuations $\mathbf{\Delta V}$. Specifically, if a value $V_m$ has been obtained after averaging on $m$ samples, one has\footnote{Note added: the following equation is imprecise for $d>2$: see a better discussion in L. Sheridan and V. Scarani, Phys. Rev. A \textbf{82}, 030301(R) (2010).}
\ba
\Delta V&=&\sqrt{\frac{\ln(1/\epsilon_{PE})+d\ln(m+1)}{2m}}\,.\label{deltav}
\ea
with $\epsilon_{PE}$ the probability that the fluctuation is actually larger than $\Delta V$ and $d$ the minimal number of outcomes of the POVM that is used to estimate $V$ (typically, $d=2$ for errors on bits: one has to estimate the probabilities of the event ``bits equal'' and of the event ``bits different'').

\end{enumerate}

Putting everything together, one obtains (\ref{finitebound}) with \ba\Delta(n)&=&\frac{2}{n}\log\frac{1}{\epsilon_{PA}}+7\sqrt{\frac{\log(2/\bar{\epsilon})}{n}}\,. \label{deltan}\ea Since failure probabilities add, the total error is
\ba
\epsilon&=& \epsilon_{PA}+\bar{\epsilon}+n_{PE}\epsilon_{PE}+\epsilon_{EC}
\ea with $n_{PE}$ the number of parameters to be estimated.

This concludes our overview of the meaning and derivation of the finite-key bound (\ref{finitebound}). As we mentioned, there are alternative approaches \cite{hay2,hay3,hay4,fung}: as they stand, these approaches are specifically tailored for the BB84 protocol; but there is no reason to doubt that they can be adapted to other protocols as well.

\section{Rapid estimates of finite key effects}

To date, the bound (\ref{finitebound}) has been applied to the BB84 protocol, both in the ideal single photon case \cite{scaren1} and in more realistic scenarios \cite{caisca}, as well as to ideal implementations of the six-state protocol \cite{scaren1} and of a modified Ekert protocol \cite{scaren2}. In spite of the differences and for a wide range of reasonable values of parameters, a common feature is observed: $r_N$ becomes larger than 0 only for $N\approx 10^5-10^6$. Here, we show with a simple estimate that this is indeed the case.

We suppose that all epsilons are of the order $10^{-3}\approx 2^{-10}$. This is quite a poor requirement: it means that the key exchange may fail once every thousand runs only; but anyway, one would only get a small overhead by choosing $10^{-9}$ instead, because the dependence in the epsilons is logarithmic.

Another useful approximation is $m\approx n\approx N/2$. Close to the critical point where $r_N$ passes from being 0 to be positive, this is always observed in exact numerical estimates of (\ref{finitebound}), and it is quite reasonable: in the critical regime, one tries to devote enough many signals to the parameter estimation, without compromising the key.

Under these two assumption, we have the following approximate values for (\ref{deltan}) and (\ref{deltav}) respectively: \ba
\Delta(n)\approx \frac{40}{N}+7\sqrt{\frac{12}{N}} &\textrm{ , }& \Delta V\approx\sqrt{\frac{9+2\ln(N)}{N}}\,.
\ea Let's consider two case studies:
\begin{enumerate}
\item \textbf{Case study: effect of $\Delta(n)$}. Suppose we are in a regime of parameters such that $r_\infty=0.1$. Neglecting the parameter fluctuations $\Delta V$, from (\ref{finitebound}) one has $r_N\approx r_\infty-\Delta(n)$. This quantity is positive for $N\approx 10^5$. Of course, the estimate becomes worse, the smaller $r_\infty$ is (for instance, as a function of the distance in a practical implementation). In summary, the finite-sample corrections to privacy amplification force a lower bound of $N\approx 10^5$ signals per run.
\item \textbf{Case study: effect of $\Delta V$}. Consider $V$ as the most typical parameter in QKD, namely an error rate. Because of the quality of optical setups, errors normally range around $1-2\%$. At which point $r_N$ is zero depends on many variables and of course on the protocol, so we cannot be very specific here. But if one requests that error rates are known to a precision $\Delta V\approx 0.5\%$, one needs $N\approx 10^6$ signals per run.
\end{enumerate}

From these case studies, we see how the estimate $N> 10^5-10^6$ arise from quite trivial estimates of the finite-key effects, independently of the details of the protocols.

\section{Conclusion}

Claims of security of QKD could not be complete without integrating the effects of finite sampling, or finite-key in short. Thanks to the work of several authors, this task has now been accomplished. One of the most striking features is the fact that, to have any final key whatsoever, approximately one million signals must be processed in each run --- and the more, the better. In this text, after reviewing the main ingredients of the finite-key bounds, I have stressed that the lower bounds on the number of signals can be obtained from very simple estimates and are therefore rather robust. While the performances of QKD may still be improved in many aspects, it seems unavoidable that QKD will always be a \textit{million-signal task}; or, to put it more positively, I would consider it as a milestone if someone could find a way of improving significantly on these bounds.

\section*{Acknowledgements}

I thank all the participants to the workshop ``Quantum cryptography with finite resources'' (Singapore, 4-6 December 2008) for very valuable comments and queries. This work was supported by the National Research Foundation and the Ministry of Education, Singapore.

\end{document}